\documentclass[aps, reprint]{revtex4-2}

\usepackage{graphicx}
\usepackage{tikz}
\usepackage{amsmath}	
\usepackage{amssymb}
\usepackage{amsfonts}
\usepackage{mathrsfs}
\usepackage{mathtools}
\usepackage{comment}

\usepackage{color}
\usepackage[]{xcolor}

\definecolor{unamblue}{cmyk}{1 0.79 0.12 0.59}

\usepackage[]{hyperref}
\hypersetup{
	colorlinks=true,%
	citecolor=unamblue,%
	filecolor=unamblue,%
	linkcolor=unamblue,%
	urlcolor=unamblue,
	bookmarksnumbered=true,     
	bookmarksopen=true,         
	bookmarksopenlevel=1,       
	pdfstartview=Fit,           
	pdfpagemode=UseOutlines,
	pdfpagelayout=TwoPageRight
}
\usepackage{caption}

\usepackage{subcaption}

\newcommand{\sfA}{\mathsf{A}}
\newcommand{\sfT}{\mathsf{T}}
\newcommand{\sfP}{\mathsf{P}}

\newcommand{\dd}{\mathrm{d}} 
 
\begin{document}
	\title{Polytope symmetries of Feynman integrals}
	\author{Leonardo de la Cruz}
	\email{leonardo.delacruz@ipht.fr}
	\affiliation{Institut de Physique Th\'eorique,  CEA, CNRS, Universit\'e Paris–Saclay, F-91191, Gif-sur-Yvette cedex, France}
%
	
	\begin{abstract}
		 Feynman integrals appropriately generalized are  $\mathsf A$-hypergeometric functions.     
		 Among the properties of $\mathsf A$-hypergeometric functions are symmetries associated with the Newton polytope. In ordinary hypergeometric functions these symmetries lead to linear  transformations. Combining tools of $\mathsf A$-hypergeometric systems and the computation of symmetries of polytopes, we consider  the associated symmetries of  Feynman integrals in  the Lee-Pomeransky representation. We compute the symmetries of $\mathtt n$-gon integrals up to $\mathtt n=8$, massive banana integrals up to 5-loop, and on-shell ladders up to 3-loop. We apply these
		 symmetries to study finite on-shell ladder integrals up to 3-loop.
	\end{abstract}
	
	\maketitle
	
	\section{Introduction}
	The evaluation of Feynman integrals is a fundamental problem in perturbative quantum field theory that is deeply connected with modern methods in mathematics \cite{Weinzierl:2022eaz}. 
	Among those is the recognition that they can be evaluated in terms of  generalized hypergeometric series known as $\sfA$-hypergeometric functions, which were proposed by  Gel'fand-Kapranov-Zelevinsky(GKZ) as a unified approach to hypergeometric functions~\cite{GELFAND1990255}.  The evaluation of Feynman integrals using GKZ is done by restricting the general solution of the system of differential equations that a Feynman integral satisfies when the coefficients of the Symanzik polynomials are promoted to be undetermined \cite{delaCruz:2019skx, Klausen:2019hrg} . The evaluation can  be done using two equivalent approaches, namely by constructing triangulations of the Newton polytope of the sum of the Symanzik polynomials or by the Gr\"obner deformation approach \cite{sturmfels:1999}. These methods have been recently automated in a package \cite{Ananthanarayan:2022ntm,Ananthanarayan:2021yar}.
	
	The matrix of exponents of the sum of Symanzik polynomials, encoded in a matrix $\sfA$,  is key in various aspects of the approach because its convex hull defines the Newton polytope.  
	Its (regular) triangulations give valid solutions to the system of differential equations and, in the case where the parameters are generic,  its volume determines the rank of the system of differential equations. Moreover, the Newton polytope determines the convergence domain of a generic Feynman integral \cite{2011arXiv1103.6273B}. It is  also  useful in numerical evaluations of Feynman integrals 
	\cite{Borowka:2015mxa,Borinsky:2020rqs, Borinsky:2023jdv} and Landau analysis  \cite{Klausen:2021yrt, Gardi:2022khw, Arkani-Hamed:2022cqe}. 
	 	
	In this paper, we will apply the GKZ 
	approach to study permutation symmetries of the Newton polytope of Feynman integrals  encoded in  $\sfA$.  Permutation symmetries and the conditions in which they hold were studied  for $\sfA$-hypergeometric functions by Forsg{\aa}rd-Matusevich-Sobieska  in Ref.\cite{JensForsgaard2019}. They  correspond to linear transformations of hypergeometric functions \footnote{For instance 
	the Pfaff transformation of the Gau{\ss } hypergeometric function ${}_2F_1$ can be understood as a result of a permutation symmetry of the integral representation of 
	${}_2F_1$.}. 
As we will see,	permutation symmetries  may in general shift the kinematic dependence from the second Symanzik polynomial to the first and vice versa, and may also permute the kinematic dependence of it between the monomials. This change is compensated by a modification of the powers of the propagators.

	The remainder of this paper is organized as follows. In Section \ref{generalized-FI}, we review generalized Feynman integrals. In Section \ref{methods},  we present the mathematical methods and give a simple example of their applications.
	In Section \ref{results}, we present our results and apply them  in Section \ref{application} to the case of  on-shell ladder integrals.
	Our conclusions are presented in Section \ref{conclusion}.
	
	\section{Generalized Feynman integrals}
	\label{generalized-FI}
	We will  consider a  Feynman  integral in Euclidean space in dimensional regularization. 
	A $L$-loop integral with $N$ propagators and $E$ independent external momenta may be written
	as
	\begin{align}
		I_F(\alpha)= \int\limits_{\mathbb{R}^L} \left(\prod\limits_{i=1}^L \frac{\dd^d k_i}{\pi^{d/2}}\right)
		\frac{1}{D_{1}^{\alpha_1} \cdots D_{N}^{\alpha_N}}, \label{FeynmanIntegral} 
	\end{align}
 where  the denominators have the form 
	\begin{align}
		D_i= (\mathsf M_i)^{rs}k_r\cdot k_s + 2 (\mathsf Q_i)^{rs} k_r\cdot p_s+ \mathsf J_i \, .  
	\end{align}
	\noindent The matrices, $\mathsf M_i$, $\mathsf Q_i$, and $\mathsf J_i$ have dimensions $L\times L$, $L\times E$, and 
	$1\times 1$, respectively. The exponents have been collected into the vector $\alpha=(\alpha_1, \dots, \alpha_N)$. 	
	 The parametric representations of Eq.\eqref{FeynmanIntegral} are expressed in terms of the so-called Symanzik polynomials  $\mathcal U$ and $\mathcal F$. 
	These can be computed  from the matrices
	\begin{align}
		\mathsf M^{rs}= \sum\limits_{i=1}^N z_i \mathsf M_i^{rs}, \  Q^{r}= \sum\limits_{i=1}^N z_i \mathsf Q_i^{rs}p_s,\  
		\mathsf J=\sum\limits_{i=1}^N z_i J_i, 
	\end{align}
	\noindent so the Symanzik polynomials are 
	\begin{align}
		\mathcal{U}= \det(\mathsf M), \quad \mathcal{F}=\det(\mathsf M)\left(\mathsf J-\left(\mathsf M^{-1}\right)^{ij} \mathsf Q^{i}\cdot \mathsf Q^{j} \right). 
	\end{align}
	\noindent
	The polynomial $\mathcal{U}$ is a homogeneous polynomial in $z$ of degree $L$, while  $\mathcal{F}$
	is homogeneous of degree $L+1$ \footnote{$\mathcal{F}$ depends on the kinematics and thus to combine it with $\mathcal U$ it must be appropriately scaled to make it dimensionless. We will leave this scaling implicit.}.
	In Euclidean
	kinematics,  $\mathcal{U}$, $\mathcal{F}$ are positive semi-definite functions of the Feynman parameters.
	These polynomials can also be obtained from the topology of the graphs. Their properties are 
	summarized in Refs.\cite{Bogner:2007cr, Bogner:2010kv}. The Lee-Pomeransky representation 
	 is given by \cite{Lee:2013hzt} 
	\begin{align}
		I_{\text{LP}}(\alpha)=\xi_{\Gamma} 
		\int_{\mathbb{R}_+^N} z^\alpha \mathcal G(z)^{-d/2} \dd \eta_N ,\  \label{Leerep}  \dd \eta_N:=\frac{\dd z_1}{z_1} \dots \frac{\dd z_N}{z_N}\,, 				
	\end{align} 
	\noindent  where we have used the multi-index notation and $\mathbb{R}_+=(0,\infty)$. The overall factor and the
	polynomial $\mathcal G (z)$ are
	\begin{align}
		\xi_{\Gamma}:=& \frac{\Gamma(d/2)}{\Gamma((L+1)d/2- \alpha_{12\dots N}) \Pi_{i=1}^N \Gamma(\alpha_i)} \, ,\\
		\mathcal G(z):=&\mathcal{U}+\mathcal{F} \, ,
	\end{align}
where $x_{ij\dots}:=x_i+x_j+\dots$ From now on, we will focus on the integral stripped from the  kinematic-independent gamma factors, namely $	I_{\mathcal G}(\alpha, \beta):=I_{\text{LP}}(\alpha)/\xi_{\Gamma_{\alpha}}$ with $\beta:=d/2$. 
	The polynomial $\mathcal G(z)$ may be written as follows
	\begin{align}
		\mathcal G(z)= \sum_i c_{i}z^{a_i},
		\label{uplusf}
	\end{align}
	where the sum runs over all monomials in $\mathcal G$ and $a_i \in \mathbb N^N $ are the exponent vectors of each monomial. The kinematic content of the integrals is  in the coefficients  $c_i$. Let us denote by $n$ the total number of monomials and consider the $(N+1) \times n$ matrix of exponents of $\mathcal G$, which we write as
	\begin{align}
		\mathsf{A}=
		\begin{pmatrix}
			1&1&\dots& 1\\
			a_1 & a_2 & \dots & a_n
		\end{pmatrix}  \, .
		\label{config-gr}
	\end{align}
The corresponding coefficients can be grouped in a vector $c=(c_1, \dots, c_n)$. 
	The Newton polytope of $\text{Newton}(\sfA)=\text{conv}(\sfA)$ is defined as the convex hull of the columns of the matrix $\sfA$. 
	 Promoting the coefficients in the polynomial $\mathcal G (z)$ to be indeterminate leads to the \emph{generalized Feynman integral} \cite{delaCruz:2019skx, Klausen:2019hrg}
	\begin{align}
		I_{\mathcal G}(\kappa , c):=
		\int_{\Omega} z^\alpha\mathcal G (c, z)^{-d/2}  \dd \eta_N,
		\label{Euler}
	\end{align}
	where $\kappa:=-(d/2, \alpha)$ and $\Omega$ is some cycle of integration. The coefficients $c$ in Eq.\eqref{uplusf} are now variables so we have indicated explicitly it writing $\mathcal G(c, z)$.  
	The first row in \eqref{config-gr} represents an overall factor $z_0$ in $\mathcal G$, whose integral is taken over $S^1$ and set to unity. 
		The integral \eqref{Euler} is a solution of  the  $\sfA$-hypergeometric  system of  differential equations:
	\begin{align}
		\Big(\partial^u -\partial^v\Big)
		F(\kappa , c)=0&, \ \text{where} \  \mathsf{A}u=\mathsf{A}v,\label{PDE1} \\ 
		\Big(\sum\limits_{j=1}^n a_{ij} \theta_j-\kappa_i\Big)F(\kappa , c)=0,& \quad 
		i=1, \dots, N+1, 
		\label{PDE2}
	\end{align}
where $\theta_j:=c_j \partial/{\partial c_j}$.  This system is usually denoted by $H_{\sfA}(\kappa)$. 
 For 
	generic parameters $\kappa$, the rank of the system satisfies the inequality \cite{sturmfels:1999} 
	\begin{align}
		\text{rank}(H_\mathsf{A}(\kappa)) \ge \text{Vol}(\text{Newton}(\sfA)),
	\end{align}
where $\text{Vol}(\text{Newton}(\sfA))$ is the normalized volume of the Newton polytope \footnote{for our case this is simply $\text{Vol}(\text{Newton}(\sfA))=N! \text{vol} (\text{Newton}(\sfA)$)}.  
	The Euler integral \eqref{Euler} is a special case of  the integrals studied by GKZ \cite{GELFAND1990255}. More generally, Feynman integrals in parametric representation are special instances of the so called Euler-Mellin integrals and their transition to $\sfA$ hypergeometric functions was discussed in Ref.\cite{2011arXiv1103.6273B}, where  non-compact cycles of integration were constructed. These are the appropriate cycles to study Feynman integrals. On the other hand, compact cycles are useful to study maximal cuts \cite{Vanhove:2018mto}. The evaluation of these integrals is done using  series solutions of the system of PDEs upon  restriction \cite{delaCruz:2019skx, Klausen:2019hrg}, see also \cite{Henn:2023tbo}.  
	\section{Mathematical methods}
	\label{methods}
	We are interested in  symmetries of a Feynman integrals, which encode symmetries of the lattice polytope $\text{Newton} (\sfA)$.  
	 Determining the symmetries of  polytopes is a fundamental problem in polyhedral computations, see e.g, the review \cite{Bremneretal2014}.
	\subsection{Normal form of a polytope}
The information about the monomials present in $\mathcal G$ is recorded in $\sfA$ so the order of the columns is  irrelevant as it defines the same system of PDEs. 
 However,  for computations a convenient way of deducing the symmetries of a polytope is to  choose a special ordering known as the \emph{normal form} of the polytope.  We say that $\sfA$ is determined only up to a $S_{n}\times \text{GL}(N, \mathbb Z)$ symmetry, where $S_{n}$ is the symmetry group  of permutations of the vertices and $\text{GL}(N, \mathbb Z)$ is the group of coordinate transformations of a $N$-dimensional lattice \cite{Kreuzer:1998vb}. An algorithm to obtain the normal 
form was used in Refs.\cite{Kreuzer:1998vb,Kreuzer:2000xy} to classify reflexive polyhedra in 3 and 4 dimensions and later implemented in PALP \cite{Kreuzer:2002uu}. A detailed description of the (PALP) normal form that we will use in this work can be found in Appendix B of Ref.\cite{grinis2013normal}. 
	\subsection{Two results by   Forsg{\aa}rd-Matusevich-Sobieska (FMS)}
Let $\phi$ be a monomial automorphism associated with $\sfA$ of the form $\phi(z_i)=z^{t_i}$ for $i=1, \dots, N+1$, where $t_i\in \mathbb Z^{N+1}$. Let $\sfT$ denote the matrix whose columns are the exponent vectors $t_i$, $\sfT= (t_1, \dots, t_{N+1})$. Applied on the monomials of $\mathcal G(c, z)$,   $\phi$ induces a permutation on $\sfA$ so that $\sfT\sfA = \sfA \sfP$, which encodes a polytope symmetry of $\text{Newton}(\sfA)$. The following statements are proved in Ref.\cite{JensForsgaard2019}: 
\begin{description}
	\item [Corollary 4.1] The monomial homogeneous automorphism $\phi$ induces the transformation 
	\begin{equation}
		F(\kappa, c) = \det(\sfT ) F(\sfT\kappa, c \sfP), 
		\label{firststatement}
		\end{equation}
	 where $c \sfP$ and $\sfT\kappa$ denote the standard matrix multiplication.
	\item [Theorem 4.4] Let $F(\kappa, c)$ be an $\sfA$-hypergeometric function for which there exists a transformation valid for generic parameters $\kappa$;
	\begin{align}
		F(\kappa, c) = R(\kappa) F(\sfT \kappa,  c \sfP),
		\label{secondstatement}
	\end{align}
	where $P$ is a permutation matrix and $R(\kappa)$ is a constant with respect to $c$. Then $\sfT \sfA=\sfA \sfP$. That is, $P$ encodes a polytope symmetry of $\sfA$ 
	\end{description}
A few comments are in order. The second statement is only a partial converse of the first one since not all polytope symmetries induce an automorphism of 
$\sfA$. For Feynman integrals the factors in Eq.\eqref{firststatement}-\eqref{secondstatement} are unity. The cycle of integration of the Euler-type integral $F(\kappa, c)$ is $\mathcal C= \mathbb R_+^{N} \times S^1$ and they are invariant under $\phi$ up to homotopy.

 The calculation of the full set of symmetries of a general Feynman integral is of course a daunting task. Permutation symmetries  grow factorially with the number of monomials in the Symanzik polynomials.  The most general application of  FMS is  to answer the question of whether certain permutation is a symmetry or not. A direct search can be done by choosing some permutation matrix, say $\sfP_{\text{try}}$, and solve a linear system for $\sfT$, namely 
\begin{align}
	\sfT \sfA \sfP_{\text{try}}-\sfA=0 \, .
\label{linear-equation}
\end{align}
If the system has a solution the corresponding pair is a symmetry of the polytope. The symmetry thus computed is a symmetry of the generalized Feynman integral and it is a property of the Newton polytope. In the examples below, we will restrict our direct searches  to cases with up to 11 monomials. When the number of monomials is higher we use PALP \cite{Kreuzer:2002uu, Braun:2012vh} by first computing the normal form of $\sfA$ and identifying the permutations that leave the polytope invariant. One then solves Eq.\eqref{linear-equation} for $\sfT$.
\subsection{Example} 
\begin{figure}[htb]
	\centering
	\includegraphics[scale=1.0]{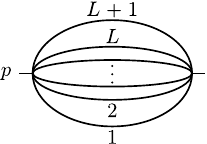}
	\caption{Banana graph: $p^2=-s$}
	\label{cantaloupe-diagram}
\end{figure}
In order to illustrate the above statements let us consider the $L=1$ bubble with a single mass (see Fig.\ref{cantaloupe-diagram}). The inverse propagators are $D_1=(k-p)^2+m^2$, $D_2=k^2$. This integral is a  Gau{\ss }  hypergeometric  function in disguise.  Its graph polynomial is represented by

\begin{align}
	\sfA=
	\begin{pmatrix}
		1 & 1 & 1 & 1 \\
		0 & 1 & 1 & 2 \\
		1 & 1 & 0 & 0 \\
	\end{pmatrix}
\end{align}
with $c=(1,  (m^2+s), 1, m^2)$ and $\sfA$ written in normal form.
Let $\kappa=-(\beta, \alpha_1, \alpha_2 )$ be the vector of generic parameters.  The calculation of the symmetries consist on finding all pairs $(\mathsf T , \mathsf P)$ such that $\mathsf T \sfA = \sfA \mathsf P$.  A direct search leads to the pairs $(\sfT_i, \sfP_i)$, $i=1, \dots, 8$,  which encode a polytope symmetry of $\text{Newton}(\sfA)$. Using the cycle notation, the permutations that lead to symmetries are 
\begin{align}
	\sigma_i= \{(),(2 3), (1 2)(3 4), (1 2 4 3), (1 3 4 2 ), (1 3) (2 4),  	\label{sigmas10} \\ (1 4), (1 4) (2 3)\} \ . \nonumber
	\end{align}
Here the empty cycle $()$ represents the permutation where all are fixed points. Their corresponding $\sfT$ matrices are
\begin{align}
		&\{\sfT_i \}= \Bigg\{
		 \mathbf{1}_3, 
		 \begin{pmatrix}
		 	1 & 0 & 0 \\
		 	0 & 1 & 0 \\
		 	2 & -1 & -1 \\
		 \end{pmatrix},
		 \begin{pmatrix}
		 	1 & 0 & 0 \\
		 	3 & -1 & -2 \\
		 	0 & 0 & 1 \\
		 \end{pmatrix}
	, 	\label{Ts10} \\
		&
		\begin{pmatrix}
			1 & 0 & 0 \\
			-1 & 1 & 2 \\
			2 & -1 & -1 \\
		\end{pmatrix}
		,
		\begin{pmatrix}
			1 & 0 & 0 \\
			3 & -1 & -2 \\
			-1 & 1 & 1 \\
		\end{pmatrix}
		, 
		\begin{pmatrix}
			1 & 0 & 0 \\
			-1 & 1 & 2 \\
			1 & 0 & -1 \\
		\end{pmatrix}
		,\nonumber\\ 
		& 
		\begin{pmatrix}
			1 & 0 & 0 \\
			2 & -1 & 0 \\
			-1 & 1 & 1 \\
		\end{pmatrix}
		, 
		\begin{pmatrix}
			1 & 0 & 0 \\
			2 & -1 & 0 \\
			1 & 0 & -1 \\
		\end{pmatrix}
	 \Bigg\}\, 
		 ,\nonumber
	\end{align}  
where $\mathsf 1_{i}$ the $i\times i$ identity matrix. 
The set of matrices form a group under matrix multiplication and thus we have included the identity matrix in the counting as we will do in the rest of the paper. The monomial automorphisms of $\sfA$ associated, say, with $\sfT_7$ are
\begin{align}
\phi(z_0)=	\frac{z_0 z_1^2}{z_2}, \ 
 \phi(z_1)= \frac {z_2}{z_1},  \
 \phi(z_2)= z_2 \ .
\end{align}
The pair $(\sfT_7, \sfP_7	 )$ encodes an identity for the generalized Feynman integral, specifically 
\begin{align}
	I_{\mathcal G }(c,\kappa)
	=	I_{\mathcal G } ((c_4,c_2,c_3,c_1), \kappa'),
\end{align} 
where $\kappa'=-(\beta,2\beta -\alpha_{1}, -\beta+\alpha_{12})$.
On the restricted integral this corresponds to the identity
\begin{align}
	I_{\mathcal G }(\kappa)=& 
	\int\limits_{\mathbb R_+^2}\dd\eta_2 
	\frac{z_1^{\alpha_1}z_2^{\alpha_2}}{(z_1+z_2+m^2 z_1^2+z_2 z_1 (m^2+s))^\beta}\\
	=&
	\int\limits_{\mathbb R_+^2} \dd\eta_2  \frac{z_1^{2\beta-\alpha_1} z_2^{-\beta+\alpha_{12}}}
	{{( z_1+m^2 z_2+z_1^2+z_1 z_2 (m^2+s))}^{\beta}}  . 
\end{align}
Notice that the transformation has moved the mass dependence to one of the monomials of $\mathcal U$ and compensated it by shifting the powers of $z_1$, $z_2$ in the integrand. 
It is easy to verify that the second integral evaluates  to 
\begin{align}
	I_{\mathcal G }(\kappa)=&
	(m^2)^{\beta-\alpha_{12}} \frac{\Gamma \left(\alpha _2\right) \Gamma \left(\beta -\alpha _2\right)
		\Gamma \left(2 \beta -\alpha _{12}\right) \Gamma \left(-\beta +\alpha _{12}\right)}{\Gamma (\beta )^2} \nonumber \\ &\times {}_2F_1\left(\alpha_2,-\beta+\alpha_{12};\beta; -s/m^2\right), 
\end{align}	 
which agrees with the result of the first integral upon using an identity of the Gau{\ss } hypergeometric function ${}_2F_1$ \cite{delaCruz:2019skx}. The matrices \eqref{Ts10} and the permutations matrices of \eqref{sigmas10} imply that there are six additional representations of the Feynman integral that can be obtained using Eq.\eqref{firststatement}. We have checked numerically that they indeed match.
\section{Results}
\label{results}
\subsection{Bananas}
\begin{table}
	\centering
	\begin{tabular}{lllll}
		\hline
		$L$ & & $[11 \dots ]$  & $[10 \dots 0 0]$ & $[10 \dots 0 1]$\\
		\hline
		$1$ & &$2!$	&   $2! \times 4 $ & $-$ \\
		\hline
		$2$ &	& $3!$  &  $3! \times 2 $ & $4$		\\ \hline
		$3$ & &  $4!^*$ &  $4! \times 2 $ & $4$ \\
		\hline
		$4$ &	& $5!^*$ &  $5! \times 2 $ &  $4^*$
	\end{tabular}
	\caption{Symmetry counting for the banana graph: the numbers include the identity matrix. Starred numbers were computed through PALP.}
	\label{results-symmetries}
\end{table}
We will consider now  the $L$-loop fully massive banana graphs (a.k.a. sunset graphs) shown in Fig.\ref{cantaloupe-diagram}. We parametrize their denominators as%
\begin{align}
	& D_1=(k_1-p)^2+m_1^2, \quad  D_2=(k_2-k_1)^2+m_2^2, \\
	& \dots, D_{L}=(k_{L}-k_{L-1})^2+m_{L}^2, \quad D_{L+1}=(k_{L})^2+m_{L+1}^2, \nonumber
\end{align}
and  label the integrals by their denominators masses.  In the above order a massless propagators is recorded as 0 and 1 otherwise. A $L$-loop banana integral is then labeled by a binary word $w$ of length $|w|=L+1$. For example, $w=10101$ labels a 4-loop integral where $D_2$ and $D_4$ are massless. The results of the calculations are summarized in Table \ref{results-symmetries}. 

Comparing the results at one loop we notice that there is only one 
symmetry for $w=11$. This corresponds to the  relabeling $z_1 \leftrightarrow z_2$, $\alpha_1 \leftrightarrow \alpha_2$. We could also interpret this relabeling as the exchange $m_1\leftrightarrow m_2 $, which leaves the polynomial invariant and it is also a symmetry of the result when expressed as a linear combination of two ${}_2F_1$ \cite{Davydychev:2000na}.
 Thus there are less symmetries than in the case $w=10$. These  are inherited from the symmetries of the Gau{\ss } hypergeometric function whose group of matrices is also generated by eight matrices.  

At higher loops we find that the only symmetries present for the massive bananas are due to relabeling. We have performed direct searches up to 2-loop for the fully massive case and up to 4-loop in the 1-mass case.  A direct search is of course prohibitive already at 3-loop for $w=1111$ where there are 17! permutation-matrices to test.  Instead, we have computed the number of symmetries with PALP and explicitly checked through Eq.\eqref{linear-equation} that the permutation matrices that lead to symmetries are relabelings.
We have checked this up to 5-loop. 

\subsection{Massless $\mathtt n$-gon}
We have also considered scalar $ \mathtt n$-point integrals up to $\mathtt n=8$. We parametrize the inverse propagators for these examples as follows
\begin{align}
	&D_1=k^2, \quad D_2=(k-p_1)^2, \\ 
	& D_3=(k-p_{12})^2, 	 \dots,   D_{\mathtt n}=(k-p_{12\dots \mathtt n-1})^2. \nonumber
\end{align}
For massless external edges we have, $p_i^2=0$.  
We have found  $72$, for  $\mathtt n=4$, and 
2 for all the other cases up to $\mathtt n=8$. The only symmetry for $\mathtt n>4$ is due to the invariance of the polynomials under the relabelings $\{z_1,z_2, \dots,z_{\mathtt n}\} \to \{z_{\mathtt n}, z_{\mathtt n-1}, \dots z_{\mathtt 1}\} $.
\subsection{On-shell ladders}
\begin{figure}[htb]
	\begin{subfigure}[b]{0.2\textwidth}
\includegraphics[scale=1]{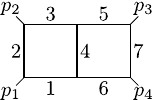}
\end{subfigure}
     \hfill
\begin{subfigure}[b]{0.2\textwidth}
	\includegraphics[scale=1]{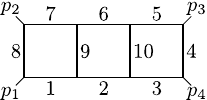}
	\end{subfigure}
	\caption{Double box (left) and triple box (right).}
	\label{on-shell-double-box}
\end{figure}
The double box graph shown in Fig.\ref{on-shell-double-box} (left) has the denominators
\begin{align}
	& D_1=k_1^2,\
	D_2=(k_1-p_1)^2,\
	D_3=(k_1-p_{12})^2,\ \\
	& D_4=(k_1-k_2)^2, \ D_5=(k_2-p_{12})^2, \  D_6=k_2^2,\ D_7=(k_2-p_{123})^2 . \nonumber
\end{align}
The matrices that represent permutation symmetries are associated to the cycles
\begin{widetext}
	\allowdisplaybreaks{
\begin{align}
	\label{cycles2box}
	&{\sigma_i}=  \\
	&	\{(3\ 4)(5\ 6)(7\ 8)(9\ 12)(10\     11)(13\ 16)(14\ 15)(17\  18)(19 \ 20)(21\ 23)(22\  24)(25\  26), \nonumber\\
	&
	(3\ 6)(4\ 5)(9\ 11)(10\ 12)(13\ 15)(14\ 16)(21\ 24)(22\ 23)(25\ 26), (3\ 4)(5\ 6)(9\ 15)(10\ 16)(11\ 13)(12\ 14)(19\ 20), (), \nonumber \\ 
	&
		(3\ 5)(4\ 6)(9\ 13)(10\ 14)(11\ 15)(12\ 16)(19\ 20)(21\ 24)(22\ 23)(25\ 26), \nonumber \\
&	(7\ 8)(9\ 14)(10\ 13)(11\ 16)(12\ 15)(17\ 18)(21\ 23)(22\ 24)(25\ 26), \nonumber \\
	&(3\ 6)(4\ 5)(7\ 8)(9\ 16)(10\ 15)(11\ 14)(12\ 13)(17\ 18)(21\ 22)(23\ 24), \nonumber
	\\
	&(3\ 5)(4\ 6)(7\ 8)(9\ 10)(11\ 12)(13\ 14)(15\ 16)(17\ 18)(19\ 20)(21\ 22)(23\ 24)
	\} \,. \nonumber
\end{align} 
}
\end{widetext}
Besides the identity, three of these symmetries
correspond to relabelings of the edge variables that leave $\mathcal G$ invariant, namely $ \{1 \leftrightarrow 3$, $5 \leftrightarrow 6 \}$, $\{1 \leftrightarrow 5, 3 \leftrightarrow 6, 
2 \leftrightarrow 7 \}$, and $\{3 \leftrightarrow 5, 1 \leftrightarrow 6, 2 \leftrightarrow 7\}$. More interesting are symmetries that shift the kinematic dependence. For instance, the first cycle gives the following representation of the Feynman integral
\begin{align}
	&I_{\mathcal{G}}(\kappa)
	= \\ 
	&\int_{\mathbb R_+^7}\dd\eta_7 \ \frac{ z_1^{-\alpha _{12}+\beta }z_2^{\alpha _2}z_3^{-\alpha _{23}+\beta }z_4^{\alpha _{1234567}-2 \beta } z_5^{-\alpha _{57}+\beta }z_6^{-\alpha _{67}}z_7^{\alpha _7}} {\mathcal G_{\sigma_1}^\beta},  \nonumber
\end{align}
where  $ \mathcal G_{\sigma_1}$ reads 
\begin{align}
	\mathcal G_{\sigma_1} (z)&=tz_2 z_4 z_7+ s z_4 \left(z_{123567}+z_6 z_{35}+z_1 z_{35}\right)
	\\
	&+ z_{23} \left(z_{567}+z_5 z_6\right)+z_1 \left(z_{567}+z_5 z_{36}+z_3 z_{67}\right), 
	\nonumber
\end{align}
where $s=-p_{12}^2$, $t=-p_{23}^2$.  This should be compared against the usual polynomial
\begin{align}
	\mathcal G_{\sigma_4} (z)&=tz_2 z_4 z_7+ s \left(z_{123}z_5z_6+ z_{1567}z_1z_3+z_{35}z_{16}z_4\right)
\nonumber	\\
	&+z_{123}z_{567}+z_4z_{123567}.
\end{align}

At 3-loop we have found that the only symmetries present correspond to edge relabelings that leave its $\mathcal G(z)$ polynomial invariant, namely $\{1\leftrightarrow 3, 5 \leftrightarrow 7, 4 \leftrightarrow 8 , 9 \leftrightarrow 10 \}$, $\{1 \leftrightarrow 7, 2 \leftrightarrow 6, 3 \leftrightarrow 5, 4 \leftrightarrow 8\}$, $\{ 1 \leftrightarrow 5, 2 \leftrightarrow 6, 3 \leftrightarrow 7, 4 \leftrightarrow 8, 9 \leftrightarrow 10 \}$. Thus, including the identity there are 4 symmetries.
\section{The evaluation of transformed Feynman integrals}
\label{application}
As an application we consider the  evaluation of the transformed ladder integrals in $d=6$ where they are finite. Let us consider first the generalized box integral 
\begin{align}
	I_{\mathcal G}(c, \kappa)=&  \int\limits_{\Omega} \frac{z_1^{\alpha_1} z_2^{\alpha_2} z_3^{\alpha_3} z_4^{\alpha_4} \dd \eta_4}
	{(c_1 z_1+c_2 z_2+c_3 z_3+c_4 z_4+c_5 z_3 z_1+c_6 z_2 z_4)^\beta} 
	.
	\label{integral-box}
\end{align}
In the restricted case
 $c=(1,1,1,1,s,t)$ and $\Omega=\mathbb R^4_+$. We find that in this case there is 
a group of 72  polytope symmetries. Let us consider for example the pair
\begin{align}
	\sfT=
	\begin{pmatrix}
		1 & 0 & 0 & 0 & 0 \\
		0 & 0 & 1 & 0 & 0 \\
		1 & -1 & -1 & 0 & 0 \\
		1 & 0 & 0 & -1 & -1 \\
		0 & 0 & 0 & 1 & 0 \\
	\end{pmatrix}, \ \sfP=
	\begin{pmatrix}
		0 & 0 & 1 & 0 & 0 & 0 \\
		0 & 0 & 0 & 0 & 1 & 0 \\
		0 & 0 & 0 & 0 & 0 & 1 \\
		0 & 1 & 0 & 0 & 0 & 0 \\
		0 & 0 & 0 & 1 & 0 & 0 \\
		1 & 0 & 0 & 0 & 0 & 0 \\
	\end{pmatrix} ,
\end{align} 
which leads to the identity 
\begin{align}
&	I_{\mathcal G}(c, \kappa)= 	I_{\mathcal G}((c_6,c_4,c_1,c_5,c_2,c_3), \kappa'
	)
	,
\end{align} 
where $\kappa'=-(\beta ,\alpha _2,-\alpha _{12}+\beta ,-\alpha _{34}+\beta ,\alpha _3)$. Restricting the integral and setting
 $d=6$, we have
 $\kappa=-(3,1,1,1)$ and $I_{\mathcal G}=I_{\mathcal G}((t,1,1,s,1,1),\kappa)$. The calculation of the remaining  cases  leads to integrals of the form 
$I_{\mathcal G}(c\sfP,\sfT\kappa)=I_{\mathcal G}(c\sfP,\kappa)$. Since the product $\sfT \kappa$ is invariant, out of the 72 symmetries there are only 4 distinct  polynomials $\mathcal G $ produced by $c \sfP$. 

The double box can be treated along the same lines. There are 8 polytope symmetries whose permutation cycles are given in  Eq.\eqref{cycles2box}. The integral converges for
\begin{align}
	\alpha_{\text{conv}}^1=&(1, 1, 1, 1, 1, 1,1) \, , \\
	\alpha_{\text{conv}}^2=&(1, 1, 1, 2, 1, 1, 1). 
\end{align}
In tune with the one loop  case, we find  
\begin{align}
	I_{\mathcal G}(c\sfP,\sfT\kappa)=I_{\mathcal G}(c\sfP,-(\beta,\alpha_{\text{conv}}^i)), \qquad i=1,2 
\end{align}
for all cases.  

We have checked numerically that the resulting integrals for the box and double box agree. At 3-loop the symmetries leave $\kappa$ invariant and thus the above property is satisfied trivially. 
\section{Conclusions}
\label{conclusion}
In this paper we have computed  the symmetries of the  Newton polytope of a Feynman integrals. On the restricted integral these symmetries represent new equivalent representations of Feynman integrals obtained by permuting coefficients in the Symanzik polynomials. We have verified up to 5-loop that the massive sunset with all masses different possesses no other polytope symmetries than those associated with relabelings.
 For ladder integrals we have used our methods in $d=6$, where they are finite. In these cases we have found that the symmetries leave the powers of the propagators invariant  $\sfT \kappa= \kappa$. It would be interesting to prove this statement in general and study Feynman integrals with numerators \cite{Gambuti:2023eqh}.
\section*{Acknowledgments}
We thank David A. Kosower, Pavel Novichkov, and Pierre Vanhove for useful discussions on related projects. We also thank  Michael Borinsky and Pierre Vanhove for comments on the manuscript.
This work is  supported by 
the European Research Council under grant ERC-AdG-885414. It has received  additional support	 by the ANR grant “SMAGP” ANR-20-CE40-0026-01. Some calculations in this paper were done with FiniteFlow \cite{Peraro:2019svx}.


\bibliographystyle{revtex4-2}	

\end{document}